# Towards Trust Proof for Secure Confidential Virtual Machines

Jingkai Mao, Haoran Zhu, Junchao Fan, Lin Li and Xiaolin Chang

*Abstract*—The Virtual Machine (VM)-based Trusted-Execution-Environment (TEE) technology, like AMD Secure-Encrypted-Virtualization (SEV), enables the establishment of Confidential VMs (CVMs) to protect data privacy. But CVM lacks ways to provide the trust proof of its running state, degrading the user confidence of using CVM. The technology of virtual Trusted Platform Module (vTPM) can be used to generate trust proof for CVM. However, the existing vTPM-based approaches have the weaknesses like lack of a well-defined root-of-trust, lack of vTPM protection, and lack of vTPM's trust proof. These weaknesses prevent the generation of the trust proof of the CVM.

This paper proposes an approach to generate the trust proof for AMD SEV-based CVM so as to ensure its security by using a secure vTPM to construct Trusted Complete Chain for the CVM (T3CVM). T3CVM consists of three components: 1) TR-Manager, as the well-defined root-of-trust, helps to build complete trust chains for CVMs; 2) CN-TPMCVM, a special CVM provides secure vTPMs; 3) CN-CDriver, an enhanced TPM driver. Our approach overcomes the weaknesses of existing approaches and enables trusted computing-based applications to run seamlessly in the trusted CVM. We perform a formal security analysis of T3CVM, and implement a prototype system to evaluate its performance.

## I. Introduction

Utilizing the capacities of public clouds can overcome the limitations of local computing power. The security of uploaded workloads becomes a concern for users because of the possible security risks of the public clouds [1]. Virtual-Machine (VM)-based Trusted-Execution-Environment (TEE) technology, like AMD Secure-Encrypted-Virtualization (SEV) [2], can eanble the construction of a Confidential VM (**CVM**) utilizing physical isolation to protect the privacy of the data in use. Currently, most major cloud vendors offer their own CVM services, e.g., Google Cloud CVM [3], Microsoft Azure CVM service [4], and Amazon AWS service [5]. However, CVMs may not ensure their security because they can be compromised by misconfiguration [6], malicious code injection [7], and so on. Users will not migrate their workloads to public clouds unless they consider the CVM to be trusted. But the trust cannot be guaranteed by the CVM itself. Therefore, how to guarantee the trust proof of a CVM is a great concern.

Typically, the user considers a CVM to be trusted if the state of the entire process from launch is as expected [8]. The state verified by the user is usually obtained by the measuring and attestation mechanism of the CVM. AMD SEV series (including SEV, SEV-ES [9], SEV-SNP [10]), ARM CCA [11], and Intel TDX [12] are all available VM-based TEE technologies, each of which provides measuring and attestation mechanisms [13]-[16]. For example, Intel TDX measures the CVM in real-time, allowing users to construct a quote to remotely verify the current state of the CVM. AMD SEV series executes initialization-time measures to obtain the state of the CVM at its launch, including the context of the memory and the CVM registers. However, all the above-mentioned technologies only measure a certain part or a certain period of a CVM. None of them can provide the proof of a CVM's state in the whole process from the initialization and booting to current state.

Trusted Platform Module (TPM)-based trusted computing [17] performs a phase-by-phase measure and extension process from a hardware-based root-of-trust (**RoT**), storing the measurements of every phase of the system in the TPM to build a chain of trust (abbreviated as **trust chain**). The measurement of each phase [18][19] in the trust chain can be used to indicate the state of this phase of a system. And the complete trust chain that starts with the RoT device can indicate the complete state of the system from booting. We usually assume the measuring approach and the measurements stored in the TPM as secure. Therefore, the complete trust chain stored in the TPM can be used as a **trust proof** in remote attestation. By verifying the trust proof, the user can determine whether the system is trusted and and secure its security. However, it is difficult for a physical TPM to meet the requirement of providing a separate TPM device for each VM in a host. This is because one host usually has only one physical TPM, but multiple VMs can be launched on it. One widely used solution is to use virtual TPM (**vTPM**) [20] as the TPM device for VMs. vTPM is a software-only implementation of the TPM solution that acts as a RoT device for VMs and provides TPM functionalities [21]. vTPM enables the existing trusted computing (TC)-based applications to work seamlessly in a VM. Using vTPM, a trust chain can be constructed for the VM as a trust proof to ensure the VM's security.

However, an insecure vTPM may threaten the measuring approach and the measurements stored in the vTPM, resulting in an insecure trust chain. Therefore, to build a secure trust chain, the security of the vTPM needs to be ensured. Protection of the vTPM focuses on protecting its runtime and ensuring the security of the Non-Volatile Random Access Memory (NVRAM) [22]. NVRAM needs to avoid privacy leakage

• The authors are with the Beijing Key Laboratory of Security and Privacy in Intelligent Transportation, Beijing Jiaotong University, Beijing, P.R.China. E-mails: {23111143, 21112051, 23111144, lilin, xlchang} @ bjtu.edu.cn.

because it contains a lot of confidential data of the vTPM, such as keys, Platform Configuration Registers (PCRs), etc. In addition, the NVRAM Binding Attacks [23] needs to be prevented because an adversary can perform NVRAM Binding Attacks to destroy the security of the vTPM by replacing the original state file with a wrong or well-constructed state file. Usually, a vTPM uses physical TPM as its RoT to ensure its security [20] [24]. In this way, the secure vTPM can build a trust chain starting from physical RoT to ensure the security of the VM.

However, the usual solutions cannot be applied in cloud confidential computing. It is because the public cloud is insecure, and the vTPM is software in the cloud lacks ways to protect its runtime memory and storage. Therefore, it is necessary to design an approach to protect the security of the vTPM and then build a trust chain using this secure vTPM to ensure CVM's security. The authors in [25] applied the unique features of the AMD SEV-SNP to construct a trust chain by implementing a vTPM into CVMs. This approach eliminates the need to protect NVRAM by not using persistent storage, and using SEV-SNP's unique features to ensure the runtime security of the vTPM. Therefore, this approach can ensure the CVM's security based on a secure vTPM. However, SEV-based CVMs are still widely used by most production cloud providers due to the advantages of more mature technology, better compatibility, and more efficiency compared to SEV-SNP-based CVMs. The approach in [25] cannot work well in SEV-based CVMs, which limits its deployment. The authors in [26] explored a vTPM solution for AMD SEV-based CVMs. The works of [25] and [26] have at least three weaknesses in the construction of a trust chain of the CVMs as follows:

- **W1: Lack of a well-defined RoT**. This undermines the security of the trust chain constructed for the running user's *A*pplications' *CVM* (**ACVM**), thus making it difficult for remote entities to consider the ACVM to be secure by verifying its trust.
- **W2: Lack of vTPM protection.** It is reasonable to use CVM to protect the vTPM and encryp NVRAM. However, the NVRAM Binding Attacks can make the vTPM insecure, as illustrated in Fig. 1(a). The user cannot verify the trust chain provided by the insecure vTPM.
- **W3: Lack of vTPM's trust proof.** The lack of vTPM's trust proof can cause problems with malicious vTPM binding. The user cannot verify the security of the vTPM bound to the ACVM. It allows an adversary or hypervisor to connect the ACVM to a malicious vTPM. Then the security of the ACVM is compromised. Fig. 1(b) illustrates the attack of malicious vTPM.

All the above discussions motivate our work. This paper aims to generate a trust proof using the vTPM solution for the AMD SEV-based CVM to ensure its trusted, and further guarantee the CVM's security. We design an approach (denoted as **T3CVM**, *T*rusted *C*omplete *C*hain for *CVM*) that combines vTPM technology to establish a complete trust chain with well-defined RoT as a trust proof for the user's AMD SEV-based ACVM. This trust proof is stored in a secure vTPM created for this ACVM. The user can verify the trust proof to prove the trust of ACVMs in remote attestation. It also allows the existing trusted computing-based applications to seamlessly work in ACVMs. T3CVM consists of three components as follows:

1) **TR-Manager.** It is deployed in a user-trusted entity as a well-defined RoT of the CVMs. The use of this component overcomes **W1**.

2) **CN-TPMCVM**. It provides an independent secure vTPM to each ACVM. The use of this component with TR-Manager can overcome **W2** and **W3.**

3) **CN-CDriver**. It is an enhanced TPM driver to protect communications between components of the T3CVM.

These three components enables the achievement of 7 seurity goals detailed in terms of security requireemnts in Section III.B. To the best of our knowledge, we are the first to explore the construction of a complete trust chain for the AMD SEV-based ACVM. Our approach has a wide range of adaptability that can also be extended to other CVMs based on VM-based TEE technologies such as TDX, CCA, etc.

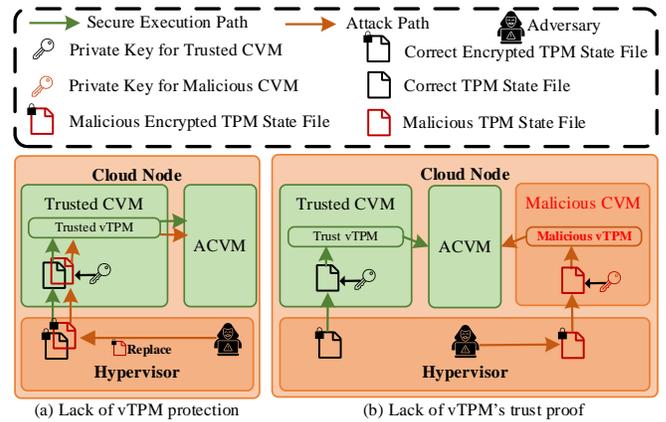

(a) Lack of vTPM protection     (b) Lack of vTPM's trust proof
**Fig. 1.** Illustration of **W2** and **W3**.

We carry out a formal analysis of the security and experimental analysis of the performance of the ACVM created by T3CVM. Prove that T3CVM can defend against the threats proposed in the threat model of Section III.A. Experiments are also conducted to evaluate the boot time and the overhead of executing TPM commands of ACVMs in T3CVM. Note that our approach is close to the work of [26]. The differences are detailed in Section II.B.

The structure of this paper is built as follows. Section II presents the background and related work, and highlights the differences between our approach and existing approaches. Section III presents the threat model and the security requirements. Section IV details the design of our approach. Section V presents the implementation of our approach. In Section VI we analyze and evaluate the security and performance of the T3CVM. Section VII concludes the paper.

## II. BACKGROUND AND RELATED WORKS

This section first presents the background in Section II.A. Then gives the related work on vTPM solutions and SEV series-based CVM's trust chain building in Section II.B.

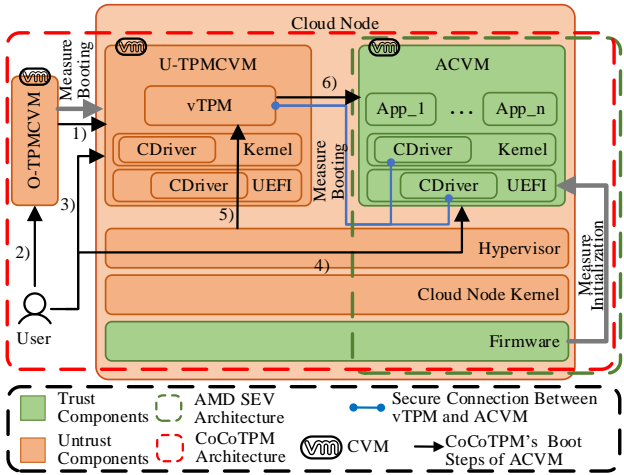

**Fig. 2.** Architecture of the AMD SEV-based ACVM and the CoCoTPM.

*A. Background*

*1) AMD SEV*

AMD SEV is a VM-based TEE technology introduced by AMD. It can create CVMs to protect the data in them. The green dashed box in Fig. 2 shows the architecture of AMD SEV-based CVM. The SEV-based CVM relies on AMD firmware, the cloud's kernel, and the hypervisor to build a VM that is isolated from the cloud's kernel, hypervisor, applications, and other VMs. The green parts in Fig. 2 represent the CVM's Trusted Computing Base (TCB), including AMD firmware and the user's ACVM (the UEFI, the kernel, and their data).

When the SEV-based CVMs launch, the memory context of the CVMs is encrypted by the firmware running on an AMD Secure Processor (AMD-SP) [27]. Then, the AMD-SP first computes the hash of the memory initialized by that CVM, which usually contains the firmware volume of the UEFI. Then, the user can obtain this measurement (called initialization measurement) for attestation. Attacks against the initialization of the CVM as well as modifications to the UEFI used are reflected in the initialization measurement. Users can know the state of the initialization phase of the SEV-based CVM by verifying the initialization measurement. Note that after the AMD firmware provides a measurement for user verification during the initialization phase, no further measurement is taken during the subsequent boot process nor the running of the CVMs.

*2) Trust Chain and vTPM*

TPM is a hardware chip including cryptographic computing components and storage components [17]. Its persistent storage NVRAM stores the keys and some privacy data. PCRs in NVRAM (namely, PCR0-PCR23) store measurements, which can build the trust chain. To address the difficulty of physical TPMs to fulfill the requirements of virtualization environments for TPM device independence, vTPM is widely used as a TPM device for VMs. vTPM is a software-implemented TPM [20] that provides a virtualized RoT for VMs. vTPM writes persistent data, such as PCRs and the user's keys to a state file. Unlike the NVRAM used by TPM chips to store measurements, vTPM stores measurements in a state file. This leads to attacks against the state file may break the trust chain. Therefore, we need to ensure that vTPM is secure. It is common practice to run it in a secure environment and to secure its communication and state files.

Building trust chains for VMs using vTPM ensures the trust of VMs to guarantee their security. The vTPM acts as the TPM device for the VM to build trust chains. The process of building a trust chain for a VM using vTPM is as follows, shown in Fig. 3. When VM boots, the Security Phase in the Virtual Basic Input/Output System (vBIOS) acts as a Core Root of Trust for Measurement (CRTM) measures the vBIOS (①) and the measurements are extended to PCR0 in vTPM (②). The measurement module in the vBIOS then measures the subsequent BootLoader (③) and extends the measurements (④). The trusted measurement module in the BootLoader measures the initialization procedures of the kernel (⑤) and extends them to the corresponding PCRs (⑥). Finally, the measurement module in the kernel measures the application in user space (⑦), extending the measurements to the corresponding PCRs (⑧). Meanwhile, vTPM runs as software in the user space of the host, which is measured by the TPM chip and ultimately anchors the trust of the VM to the CRTM in the TPM chip.

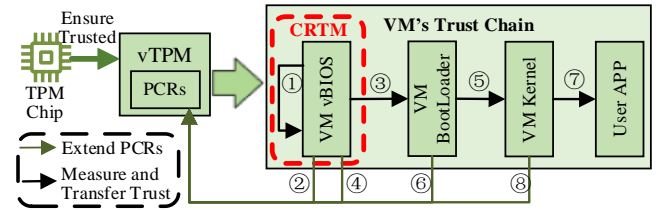

**Fig. 3.** VM's Trust Chain Built by vTPM.

*B. Related Work*

This section focuses on cloud providers' vTPM solutions for their VM services and the research on building a trust chain for SEV series-based CVM.

1) vTPM Solutions

Cloud providers usually offer vTPM solutions for their VM services as RoT to build trust chains for VMs. Google proposes Shielded VM to provide vTPM for its VM services, including measured boot and integrity monitor [28]. Microsoft Azure CVM services provide a trust proof to users through the Azure Attestation Service provided by Microsoft [29]. The Azure Attestation Platform receives evidence from CVM and provides it to the user for proof. Amazon AWS presents NitroTPM [30]. It provides TPM functionality for AWS services as well as cryptographic proof of integrity for AWS EC2 instances. However, all these solutions' security is guaranteed by the cloud providers. The user needs to trust the cloud providers, but this is prohibited in the requirements of confidential computing. Our approach allows the user to launch a vTPM in a special CVM through the RoT. This approach allows users to use a secure vTPM without having to trust the cloud provider.

2) SEV-Series-based CVM's Trust Chain Building

The authors in [31] developed a SEV-SNP-based approach to let end users carry out remote attestation of a CVM using a web browser and its extensions. However, the approach did not measure the runtime process, resulting in an incomplete trust chain. In particular, we focus on solutions that use vTPM to build a trust chain for SEV series-based CVMs. A comparison between our approach and other solutions that utilize vTPM is shown in TABLE I. Narayanan et.al. [24] explored the unique features of SEV-SNP to propose a vTPM architecture to ensure the security of a SEV-SNP-based CVM. However, the unique features of SEV-SNP make it difficult to extend the proposed architecture to SEV-based CVMs.

TABLE I
COMPARISON OF T3CVM AND EXISTING SOLUTIONS TO BUILD TRUST CHAIN FOR CVM USING vTPM

| Ref. | Root-of-Trust | Persistent state protection | NVRAM Binding Protection | Secure vTPM |
|---|---|---|---|---|
| SVSM-vTPM [25] | AMD | N/A | N/A | Yes |
| CoCoTPM [26] | Indefinite | Encrypted (In Cloud Node) | No | No |
| Our approach T3CVM | RoT Component | Encrypted (In User-Trusted Entity) | Yes | Yes |

CoCoTPM developed in [26] explored vTPM for building a trust chain for a SEV-based ACVM, illustrated in the red dashed box in Fig. 2. CoCoTPM contains two components: 1) **TPMCVM**, which is a CVM providing vTPM functionalities to the ACVM. 2) **CDriver**, which is an extended TPM driver. CoCoTPM is close to our work. The steps for users to boot their ACVMs are as follows, showed in Fig. 2, where U-TPMCVM denotes the the user's own TPMCVM, and the O-TPMCVM denotes an existing available TPMCVM as a RoT of U-TPMCVM:

1) A user utilizes an O-TPMCVM to launch a user's own U-TPMCVM.
2) The user obtains U-TPMCVM's attestation report from O-TPMCVM and authenticates it.
3) The user initializes the U-TPMCVM.
4) The user launches an ACVM, the hypervisor binds the ACVM to the user's U-TPMCVM.
5) U-TPMCVM verifies initialization measurement and the hypervisor sends the TPM state file stored in the insecure data center to it.
6) U-TPMCVM creates vTPM with the state file and continues the booting process.

The red dashed box in Fig. 4 illustrates the trust chain created by the CoCoTPM approach. This approach uses the vTPM as the TPM device to build the trust chain for an ACVM. To provide vTPM to an ACVM, a TPMCVM must first boot up and then deploy the vTPM. Thus, to ensure the security of vTPM, a trust chain must first be established for the newly booted TPMCVM. CoCoTPM assumes that there exists another TPMCVM which acts as the RoT for U-TPMCVM to build a trust chain. CoCoTPM did not describe what will be RoT of O-TPMCVM. That is, there is an incomplete trust chain in CoCoTPM.

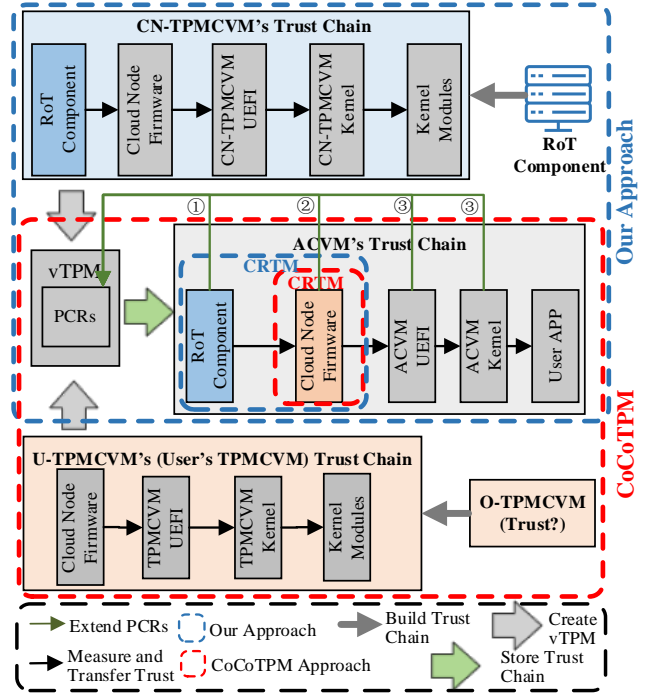

Fig. 4. Trust chain of the CoCoTPM and the T3CVM.

According to the above introduction, we find that the limitations of CoCoTPM and the major differences between our approach from CoCoTPM are as follows:

● **Build well-defined RoT.** The CoCoTPM approach deploys a user's U-TPMCVM by using O-TPMCVM as its RoT. However, there is no solution to the RoT of the O-TPMCVM. That is, the trust chain of U-TPMCVM is not complete, it cannot prove the trust of U-TPMCVM. There is a lack of the RoT for the vTPM created by this U-TPMCVM.
  **Our approach** is designed with a component (described in detail in Section IV.B) as a well-defined RoT, and it can address the weakness of **W1** (security analysis in Section VI.A). Detailed descriptions are given in Section IV.B for the RoT component description.
● **Defend against NVRAM Binding Attacks.** CoCoTPM encrypts vTPM's persistent storage using the U-TPMCVM's public key and stores it in an insecure data center. Since different state files are encrypted using the same key, it causes that an incorrect state file can be transferred to the U-TPMCVM by a malicious hypervisor in step 5) and still be correctly decrypted for use in step 6), which leads to the NVRAM Binding Attacks.
  **Our approach** can effectively mitigate NVRAM Binding Attacks, and address the weakness of **W2** (security analysis in Section VI.A). Detailed descriptions are given in Section

IV.B for component descriptions and in Section IV.C for trusted boot scheme descriptions.

- **Resist malicious vTPM**. CoCoTPM binds the ACVM to the U-TPMCVM in step 4). The U-TPMCVM will provide vTPM to this ACVM. However, this binding operation is performed by an insecure hypervisor. The ACVM may be bound to the malicious TPMCVM to use a malicious vTPM without the user being able to detect this problem. In this way, the ACVM will use a malicious vTPM.

  In contrast, **our approach** can resist malicious vTPM, it can address the weakness of **W3** (security analysis in Section VI.A). Detailed descriptions are given in Section IV.C for trusted boot scheme descriptions.

### III. THREAT MODEL AND SECURITY REQUIREMENTS

This section describes the threats occurring in the creation of vTPM and ACVM in Section III.A. Section III.B gives the security requirements based on the threats.

*A. Threat Model*

**Threats to Creating vTPM**. We consider three threats. The first is the threat against the launch process of the CVM that offers the vTPM. The CVM is deployed in the cloud node, which is insecure except for its hardware. The CVM's image includes the configuration and services that provide secure vTPMs. An adversary may modify and steal part of the CVM's image. During the launch of the CVM, an adversary may degrade its security. The second is the threat of the vTPM at runtime. The vTPM's necessary data are stored in the RoT component. An adversary may obtain or destroy sensitive information and configurations, either while they are at rest or during transmission. The third is the security of NVRAM. The adversary may steal the data in the NVRAM or perform NVRAM Binding Attacks.

**Threats to Creating ACVM**. ACVM has the following two threats. When the ACVM is launched, it needs to establish a connection with the vTPM to transmit information to build a trust chain. A privileged adversary can listen to the communication data or modify it during this phase to destroy the security of the ACVM. This is the first threat. In addition, the adversary can redirect the communication traffic to a malicious vTPM. This is the second threat.

Note that this work focuses on the threats in building a complete trust chain for AMD SEV-based ACVM with a secure vTPM. Vulnerabilities and threats to the TEE technology and the physical attacks are not considered in our threat model, e.g., ciphertext-side channel attacks [32]. Attacks against cryptographic algorithms are also not considered. The attacks on Integrity Measurement Architecture (IMA) involved in the approach, e.g., TOCTOU [33], and denial-of-service attacks (DDoS) are also out of the scope of this paper.

*B. Security Requirements*

According to the threat model given in Section III.A, we give the security requirements (SR) that T3CVM needs to defend against the threats.

**SR1: Trusted RoT Component.**

**SR2: Secure Data at Rest.**

**SR3: Secure Launch Process of vTPM Manager.**

**SR4: Secure Transmission.**

**SR5: Secure NVRAM**.

**SR6: Secure vTPM Binding**.T

**SR7: Secure Launch Process of ACVM**.

### IV. TRUSTED COMPLETE CHAIN FOR CVM (T3CVM)

In this section, we describe our approach in detail, including a description of the architecture formed by our approach (Section IV.A), the components included in our approach (Section IV.B), the CVM's trusted boot scheme in our approach (Section IV.C), and the trust chain constructed by our approach (Section IV.D).

*A. T3CVM Architecture*

Fig. 5 illustrates the architecture deploying the T3CVM approach. There are three participants:

**I) Certification Authority (CA)**, responsible for issuing certificate entities for secure communication.

**II) Cloud Node**, an AMD SEV physical machine in the public cloud that can offer CVM services to users. Each Cloud Node has applied for a certificate from CA. Cloud Node can obtain the remote entity's credentials and then establish a secure connection over the network to launch the CVM using the image that is transferred from the remote entity.

**III) User Node**, a host owned by the user who wants to deploy the secure CVM in the cloud. User Node's execution environment is assumed to be secure, and all local programs and data are well protected. The User Node can create various files and identifiers, possesses the ability to perform encryption/decryption correctly, provides networking capabilities, and ensures secure storage. Note that, to fulfill the requirements of T3CVM, we require the User Node to have a pair of root keys $TRK_{priv}$ and $TRK_{pub}$ that provide signing and other functions for the data, and the User Node should be able to communicate securely with the CA to provide certificate signing.

The blue parts in Fig. 5 denote the three components implemented in T3CVM. A brief description of the three components follows, specifically described in Section IV.B and implemented in Section V:

1) **TR-Manager**, running in a user-trusted entity, provides CN-TPMCVM with management and attestation capabilities and users with secure data storage and the user's ACVM deployment capabilities.

2) **CN-TPMCVM**, a special CVM launched by TR-Manager in the Cloud Node, provides secure vTPMs to users' ACVMs.

3) **CN-CDriver**, which is an enhanced TPM driver, runs in the UEFI and kernel of both the CN-TPMCVM and the user's ACVM.

We then present an example, shown in Fig. 5, to illustrate the architecture of the T3CVM, which includes both the components within the T3CVM and their communications. In this example, the TR-Manager, which is deployed in the User Node, launches both a CN-TPMCVM and a user's ACVM in the Cloud Node. In fact, the TR-Manager can be deployed in an entity that can be trusted by the user and can launch multiple CN-TPMCVMs and users' ACVMs. Furthermore, CN-TPMCVM can be deployed in any entity with CVM capabilities and can provide independent vTPM services to multiple users' ACVMs. The TR-Manager consists of a TR Module and two storage structures: TPM-List and VM-List. The CN-TPMCVM consists of a CN-MvTPM Module and a vTPM created for the user's ACVM. CN-CDriver acts as a TPM driver, running in the UEFI and kernel of both the CN-TPMCVM and the ACVM. Three channels are established between them for communication via TCP/IP, with the communication data being encrypted. The detailed establishment process and functions are described in Section IV.C.

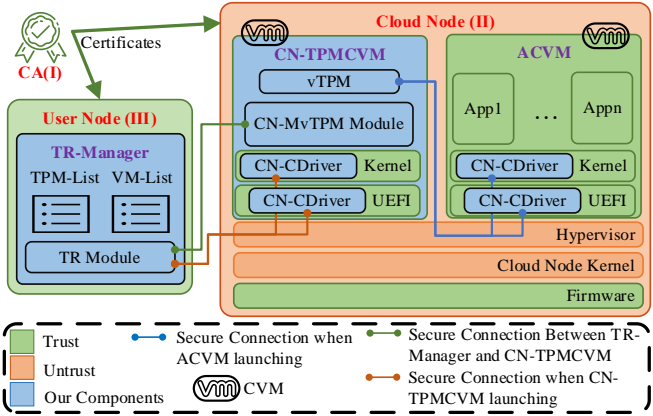

**Fig. 5.** Illustration of T3CVM Architecture.

### B. T3CVM Components

This section provides a detailed description of the components in the T3CVM approach. We will describe how to make T3CVM satisfy **SR1-SR7**.

#### 1) TR-Manager

TR-Manager is the core component of our approach. It acts as a RoT for the CN-TPMCVM and the ACVM, utilizing TCP/IP to communicate with other components. By adding this component, the T3CVM solves the **W1**. As a RoT, it needs to meet the requirements of **SR1-SR4**. We first introduce TR-Manager's functionalities in detail and then discuss how it fulfills all requirements.

**Functionalities.** The functions of TR-Manager include four aspects: initialization and authorization, secure storage, secure communications, and vTPM-like service:

1) **Initialization and Authorization.** This function is responsible for initializing and authenticating the TR-Manager in a user-trusted entity and authenticated by the user-trusted entity. For TR-Manager's initialization process, it first creates a new key pair MRKpriv and MRKpub as the root key. After this, TR-Manager initializes two special storage structures. Finally, the TR-Manager opens network ports for receiving commands. For the TR-Manager authorized by the user-trusted entity, the user-trusted entity performs static measuring execution on the TR-Manager before deploying it and stores this measurement for user attestation. Then the user-trusted entity signs the MRKpub with its private root key and requests the CA to issue a certificate.

2) **Secure Storage.** This function is responsible for the secure storage of data associated with the CN-TPMCVM and the user's ACVM. The TR-Manager consists of two storage structures, TPM-List and VM-List. The two storage structures are built in the secure storage space provided by the user-trusted entity for the TR-Manager. This space is only allowed to be used by the TR-Manager and the user, other entities are not allowed to access this space, thus ensuring the privacy of the stored data. The VM-List has many entries, each entry represents an ACVM. Within each entry, there are the following items: the *User Identifier* is the user's identification, which is used to identify which user this ACVM belongs to. The *AMD Key* is the key that is used to establish a secure channel with the AMD firmware, such as the Guest Owner Diffie-Hellman (GODH) Key Pair [34]. The *VM Key* is a unique key for every ACVM. The *ACVM Image* is the ACVM's image. The *TPM State File* is the vTPM's persistent storage. It is encrypted with the VM Key.

TPM-List also has many entries, each of which represents one CN-TPMCVM. Within each entry, there are the following items: the *CN-TPMCVM Root Key Pair* is the root key of this CN-TPMCVM. The *RK Certificate* is the certificate of the CN-TPMCVM's root key, it is signed by the user-trusted entity's root key. The *Image Key* is used to encrypt/decrypt the CN-TPMCVM's image. The *Measure Key* is a symmetric key, used to protect communication from CN-CDriver in the CN-TPMCVM to TR-Manager. The *TLS Key Pair* is used to protect the communication between the TR-Manager and CN-MvTPM Module in the CN-TPMCVM. The *Initialization Measurement* is used to store this CN-TPMCVM's initialization measurement that gets from firmware. The *Boot Measurements* are used to store this CN-TPMCVM's boot measurements. The *ACVM Pointer List* is a list of pointers. When an ACVM is bound to this CN-TPMCVM, a pointer is added to this list, corresponding to the entry in the VM-List.

3) **Secure Communications.** This function is responsible for protecting communications between the TR-Manager and other components. TR-Manager uses the current mainstream secure communication protocol, SSL/TLS protocol to protect communications. TR-Manager's communications can be divided into two parts. The first part is the communication during CN-TPMCVM is launching. In this part, the TR-Manager generates an encryption key and sends it to CN-TPMCVM when it is booting, which is used to protect CN-TPMCVM's boot measurements. The second part is the communication after the CN-TPMCVM is booted. In this part, the TR-Manager establishes a secure channel with the CN-MvTPM Module. The TR-Manager generates a TLS Key Pair and embeds it when building the image of CN-TPMCVM. When CN-TPMCVM is booted and verified by the TR-Manager, the TR-Manager constructs a secure channel with the CN-MvTPM Module.

4) **vTPM-like Service.** This function is responsible for constructing a complete trust chain for the CN-TPMCVM. This service is similar to the vTPM service but only provides services related to the construction of the trust chain such as obtaining measurements and extending it. Whenever a CN-TPMCVM is launched, the vTPM-like service is called to obtain its initialization measurement, then stores it in the TPM-List. After that, the vTPM-like service obtains the CN-TPMCVM's boot measurements and extends them into the PCRs-like space. When the CN-TPMCVM's boot process is complete, the vTPM-like service extends the measurements from PCRs-like space to the Boot Measurements item in the TPM-List.

**Solutions of Security Requirements.** With the above functions, the TR-Manager fulfills the **SR1-SR4**. For **SR1**, the TR-Manager's initialization and authorization process can ensure that it is trusted. For **SR2**, the TR-Manager has two storage structures for storing the data of users' ACVMs and CN-TPMCVMs securely. For **SR3**, the TR-Manager has a vTPM-like service to construct a complete trust chain for the CN-TPMCVM to ensure its security. In this way, secure CN-TPMCVM can create a secure vTPM. For **SR4**, the TR-Manager has a scheme for constructing a secure channel using cryptographic techniques.

*2) CN-TPMCVM*

The CN-TPMCVM is responsible for creating a vTPM for each ACVM bound to it. It manages the vTPM until the ACVM is shut down, after which the vTPM is destroyed. It communicates with other components via TCP/IP while using encryption to secure the communication. Its image is created by TR-Manager, and to minimize the attack surface, only necessary libraries and modules are kept in its image and remain unchanged after boot. By adding this component, combined with TR-Manager, the T3CVM can solve the **W2** and **W3**. As a special CVM to provide vTPM service, it needs to meet the requirements of **SR2-SR6**. We first give the design of the CN-TPMCVM, and then discuss how it fulfills all requirements.

**Functionalities.** The CN-TPMCVM contains four functionalities: encrypted image, secure communication, secure NVRAM, and secure vTPM binding:

1) **Encrypted Image.** This function is responsible for protecting the CN-TPMCVM's data at rest. The image of CN-TPMCVM is generated by TR-Manager and encrypted using Image Key. It can prevent attackers from accessing the data in this image. When CN-TPMCVM is launched, TR-Manager will securely hand over the Image Key to CN-TPMCVM to decrypt the image.

2) **Secure Communication.** This function is responsible for protecting the communication between CN-TPMCVM and TR-Manager after it has booted. The TR-Manager embeds the key used for the secure channel into the image of the CN-TPMCVM. This key will be used to encrypt the communication when the TR-Manager has verified the trust chain of the CN-TPMCVM. TR-Manager will send the required root keys, certificates, etc. to CN-TPMCVM for initialization.

3) **Secure NVRAM.** This function is responsible for protecting the NVRAM of the user's vTPM. It can protect the privacy of NVRAM and mitigate the NVRAM Binding Attacks. The NVRAMs of different ACVMs are encrypted by the unique key (VM Key) for each ACVM. Moreover, the ACVM's NVRAM and VM key are provided by the TR-Manager using a secure communication channel.

4) **Secure vTPM Binding.** This function is responsible for binding a secure vTPM with the user's ACVM and providing its trust proof. TR-Manager uses VM Key to calculate the HMAC result of the CN-TPMCVM's boot measurements and sends it to CN-TPMCVM. Then CN-TPMCVM extends the HMAC result to PCR0 of the vTPM. It can provide the vTPM's trust proof to the user's ACVM, and prevent the error binding of the user's ACVM with a malicious vTPM.

**Solutions of Security Requirements.** By the functions above, the CN-TPMCVM can fulfill the **SR2-SR6**. For **SR2**, the CN-TPMCVM applies image encryption to protect the data at rest in the image. For **SR3**, the T3CVM designs a trusted boot scheme for CN-TPMCVM to ensure the security of CN-TPMCVM, and further ensure the security of vTPM. For **SR4**, the CN-TPMCVM uses cryptographic techniques to secure the transmitted data. For **SR5**, the NVRAM is encrypted to prevent data leakage. To address the problem of NVRAM Binding Attacks, CN-TPMCVM supports a method for NVRAM to be encrypted by a unique key. For **SR6**, CN-TPMCVM supports a method for ACVM to bind with secure vTPM in CVM's trusted boot scheme.

*3) CN-CDriver*

The CN-CDriver component is an enhanced TPM driver located in the UEFI and kernel of both CN-TPMCVM and the ACVM. It is responsible for obtaining keys to encrypt and decrypt TPM commands. And, it communicates with other components via TCP/IP. As an enhanced TPM driver, it needs to fulfill the requirement of **SR7**.

**Functionalities.** CN-CDriver contains two functions, including secure communications and secure keys obtaining:

1) **Secure Communications.** This function is responsible for protecting the communication between CN-CDriver and vTPM or vTPM-like service. The CN-CDriver utilizes the communication scheme from the CoCoTPM that uses the AES-GCM algorithm to protect the TPM commands, with added support for our approach's communication protocols.

2) **Secure Keys Obtaining.** This function is responsible for obtaining the communication keys. The CN-CDriver which is located in the CN-TPMCVM can obtain the Image Key and the Measure Key provided by the TR-

Manager. The CN-CDriver which is located in the ACVM obtains the session key created by the TR-Manager.

**Solution of Security Requirement.** CN-CDriver combines with TR-Manager and CN-TPMCVM to address **SR7**. For **SR7**, the CN-CDriver supports a method to obtain keys and uses keys to protect the security of the CN-CDriver's communications. Protecting the communication ensures that vTPM-like service builds a secure trust chain for CN-TPMCVM to ensure the security of CN-TPMCVM. This ensures that the vTPM can construct a secure trust chain for the user's ACVM to ensure the security of the ACVM.

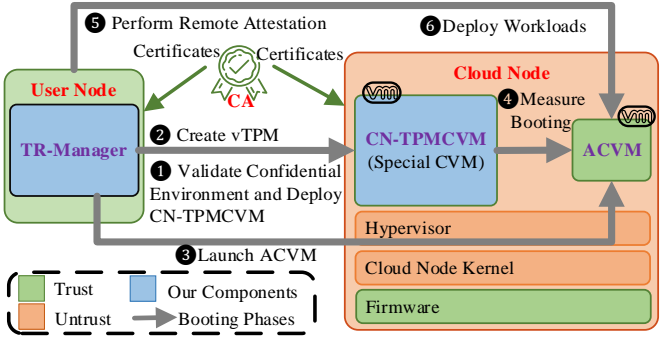

**Fig. 6.** The phases of booting an ACVM by the CVM's trusted boot scheme.

### C. CVM's Trusted Boot Scheme

In this section, we describe the CVM's trusted boot scheme in the T3CVM approach.

This scheme can launch a secure ACVM for the user. Fig. 6 shows the phases of booting an ACVM using the CVM's trusted boot scheme. The TR-Manager is deployed in a User Node which acts as a user-trusted entity, manages a CN-TPMCVM, and launches an ACVM in the Cloud Node. Note that we assume that the TR-Manager has been deployed in the User Node as a RoT. Additionally, the user has registered the ACVM in our scenario. Phases ❺ and ❻ can be realized differently in different user application scenarios and are not described in our scenario. The remaining steps are the core of the CVM's trusted boot scheme. Next, we describe the CVM's trusted boot process in detail by dividing it into two parts: the CN-TPMCVM's launch process and the ACVM's boot process.

#### 1) CN-TPMCVM's Launch Process

The phases ❶ details of the CN-TPMCVM's launch process. As shown in Fig. 7, the phase ❶ consists of 10 steps, as follows:

Step ①: TR-Manager first generates a TPM-List entry for the CN-TPMCVM. Then TR-Manager creates and stores a CN-TPMCVM Root Key Pair, signs it with the user-trusted entity's root key, and requests the CA to issue a certificate.

Step ②: TR-Manager verifies the certificate chain of the confidential environment in Cloud Node [34].

Step ③: TR-Manager generates and stores the TLS Key Pair and the Measure Key, and requests CA to issue a certificate.

Step ④: TR-Manager creates a CN-TPMCVM's minimized image and embeds the TLS Key Pair and the certificate in it. Then, the TR-Manager creates an Image Key to encrypt this image. Then, the TR-Manager sends the image to the Cloud Node to launch the CN-TPMCVM.

Step ⑤: TR-Manager obtains the initialization measurement of CN-TPMCVM, validates it, and stores it in the entry.

Step ⑥: TR-Manager sends the Image Key and Measure Key to CN-TPMCVM via AMD secret injection scheme [27].

Step ⑦: CN-TPMCVM decrypts the image and continues the boot process. The CN-CDriver continuously measures booting and uses the Measure Key to encrypt and transmit it to the TR-Manager.

Step ⑧: TR-Manager first stores the initialization measurement in PCRs-like space, then continuously extends the boot measurements. After booting, the TR-Manager validates them and stores the final boot measurements in the item Boot Measurements of the entry.

Step ⑨: TR-Manager enables the TLS Key Pair and establishes a secure communication channel with the CN-MvTPM Module in CN-TPMCVM.

Step ⑩: TR-Manager deploys the CN-TPMCVM Root Key Pair and the RK Certificate to CN-TPMCVM.

#### 2) ACVM's launch Process

Phases ❷ - ❹ are the ACVM's launch process phases.

Phase ❷ is responsible for creating a secure vTPM for the user's ACVM.

The CN-TPMCVM does not hold any persistent data, so it establishes a secure channel with the TR-Manager to obtain the necessary data. All the data it needs is provided by the TR-Manager over the secure channel, e.g., when a pre-existing user's ACVM wants to launch, the TR-Manager will send the TPM state file and the VM Key to CN-TPMCVM.

If someone wants to create a new vTPM, the TR-Manager performs the following steps to complete the creation of a vTPM. First, CN-TPMCVM creates a new endorsement key pair and a certificate. Then CN-TPMCVM signs the certificate with the CN-TPMCVM Root Key. After that, the vTPM initializes a counter with 0 to mitigate TPM Rollback Attacks [23]. Finally, this vTPM completes the initialization process.

Phase ❸ is responsible for the trusted launch of the user's ACVM. Phase ❹ is responsible for measuring ACVM's boot process using vTPM. Fig. 8 illustrates the ACVM's launch process. It consists of 10 steps, as follows:

Step ①: The launch request is sent to the TR-Manager.

Step ②: TR-Manager obtains the ACVM's image from the VM-List and sends it to the hypervisor to launch an ACVM.

Step ③: TR-Manager obtains the initialization measurement of the ACVM and verifies it.

Step ④: TR-Manager obtains the VM Key and the TPM state file from the VM-List and sends it to CN-TPMCVM using the secure channel.

Step ⑤: TR-Manager generates a session key between the vTPM and the CN-CDriver for this launch process. Then, the key is sent to the vTPM and the ACVM through two secure channels.

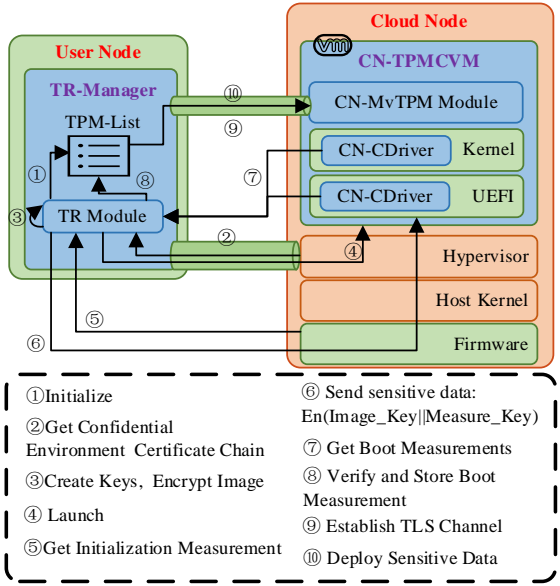

**Fig. 7.** CN-TPMCVM's Launch Process.

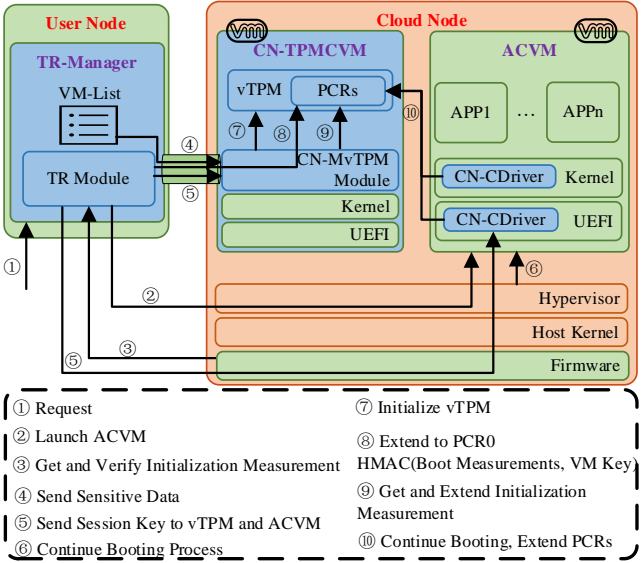

**Fig. 8.** ACVM's Launch Process.

Step ⑥: Continue the ACVM's booting process.

Step ⑦: CN-TPMCVM decrypts the TPM state file using the VM Key and delivers the decrypted file to the vTPM that has been initialized in Phase ❷.

Step ⑧: CN-TPMCVM requests the TR-Manager to calculate the HMAC result about the CN-TPMCVM's boot measurements with the VM Key, and extends this result to PCR0 of the vTPM.

Step ⑨: vTPM extends the initialization measurement to PCR0.

Step ⑩: The launch process continues and the boot measurements are extended to the corresponding PCRs.

*D. ACVM's Trust Chain*

In this section, we detail the complete trust chain constructed by our approach for the user's ACVM to prove the trust of the ACVM.

In T3CVM, we use the TR-Manager as a RoT to construct a trust chain for the user's ACVM and transfer trust from the user-trusted entity to the ACVM. We describe in detail the transfer of trust and the establishment of the trust chain below.

**Trust Transfer to the TR-Manager**. The trust transfer from the user-trusted entity (i.e., the User Node) to the TR-Manager, is mainly done by the user deploying the TR-Manager on the User Node. The User Node's root key signs the TR-Manager's root key and requests its CA to generate a certificate. In this way, the user can trust the TR-Manager that has been authenticated by the User Node.

**Trust Transfer to the CN-TPMCVM**. The CN-TPMCVM implementation process can be divided into two parts: launching a CVM and running the CN-TPMCVM's application. T3CVM provides a scheme (described in Section IV.C) to let TR-Manager build a complete trust chain for CN-TPMCVM and secure its communications. First, the TR-Manager verifies the confidential computing environment of the Cloud Node, and adds the firmware to the trust chain. Then, the TR-Manager gets and verifies the initialization measurement, and then adds the CN-TPMCVM's UEFI to the trust chain. Finally, the vTPM-like service performs the measured boot for the CN-TPMCVM and adds the kernel to the trust chain. We do not perform continuous measures at the CN-TPMCVM's runtime. This is because CN-TPMCVM runs with a minimal image and the image does not change during runtime.

**Trust Transfer to the vTPM**. The user-trusted CN-TPMCVM can communicate with the TR-Manager over a secure channel and create a vTPM for the user's ACVM. Since the user trusts the TR-Manager and the CN-TPMCVM, and the TR-Manager communicates with the CN-TPMCVM over a secure channel, the user also trusts the vTPM. As a result, the user's trust is transferred to the vTPM.

**Trust Transfer to the ACVM**. For trust transfer to the ACVM, the user performs a measured boot for the ACVM by an already secure vTPM to transfer the trust to the ACVM. The trust chain for ACVM is shown in the blue dashed box in Fig. 4. The trust transfer process is as follows:

Step ①: TR-Manager uses VM Key to calculate the HMAC result of the CN-TPMCVM's boot measurements and extend this result into vTPM's PCR0.

Step ②: CN-TPMCVM gets the initialization measurement of the ACVM and extends it to the PCR0, thus adding the ACVM's UEFI to the trust chain.

Step ③: The ACVM continues to perform the measured boot to add the kernel and each application to the trust chain.

Thus, the trust chain built for the user's ACVM is completed and the user transfers his trust to the ACVM. The TR-Manager and Cloud Node firmware together act as the CRTM for the user's ACVM, as shown in the blue dashed box in Fig. 4. The chain starts from a trusted RoT (i.e., the TR-Manager) and includes a secure vTPM, as well as the user's ACVM.

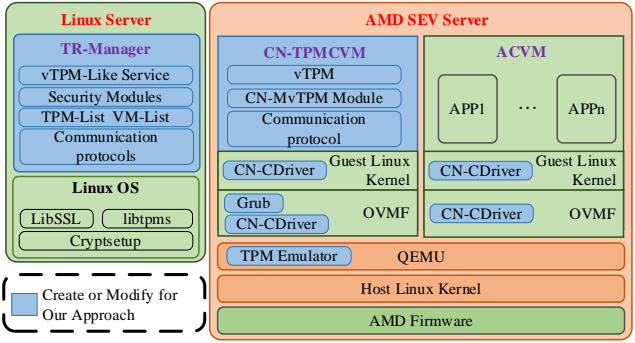

**Fig. 9.** Software stack of our implementation.

## V. IMPLEMENTATION

We implemented a prototype system of our concept, shown in Fig. 9. The system consists of a TR-Manager as a service running in a Linux Server, a CN-TPMCVM and an ACVM implemented as the AMD SEV-based CVMs running in an AMD SEV Server.

The Linux Server acts as the User Node, providing a runtime environment for TR-Manager and ensuring its security. Both CN-TPMCVM and ACVM are hosted on the AMD SEV Server. The software stack on the AMD SEV Server is recommended by AMD in GitHub [35] and we modify them to support our system. It consists of three parts, first is the QEMU, a very popular VM emulator. We modify the TPM Emulator in QEMU to support forwarding the encrypted CN-TPMCVM's measurements and the encrypted TPM commands. The second and third parts are the Open Virtual Machine Firmware (OVMF) and Linux kernel, where the OVMF serves as the firmware for the CVMs and the Linux kernel serves as the operating system kernel for the CVMs. We modify the OVMF and the Linux kernel, add the CN-CDriver in them to support our communication protocols.

### A. TR-Manager

The TR-Manager is a service in the AMD SEV Server. TR-Manager utilizes the filesystem and components in the Linux Server such as `OpenSSL` [36], `Cryptsetup` [37], `libtpms` [38], etc., and rewrites some of the `SWTPM`'s interfaces [39] to implement the functionalities below. It first generates its own root key and calls the Linux Server's root key to sign and issue the certificate. Then TR-Manager constructs two data lists, TPM-List and VM-List, using linked list and local storage structures. TR-Manager implements a security module, the functions include generating keys and certificates, and establishing secure communication channels. The `Cryptsetup` and a minimized Linux image are used to generate an encrypted image of the CN-TPMCVM. Using `libtpms` and the rewritten `SWTPM`'s interfaces, we build a vTPM-like service. And, the TR-Manager establishes a QMP port and a TCP port. These ports are used to communicate with CN-TPMCVM.

### B. CN-TPMCVM

We implement the CN-TPMCVM in the AMD SEV Server. It uses our modified OVMF and Linux kernel. In OVMF, our patch has about 410 LoC used to implement the communication protocol between TR-Manager and CN-TPMCVM. Moreover, similarly to the scheme [40], we implement the Grub as the BootLoader in OVMF to decrypt the image of CN-TPMCVM. In the Linux kernel, our patch has about 260 LoC to implement the functions including getting the Measure Key, and encrypting/decrypting the measurements of each part of the kernel.

In the CN-TPMCVM, we utilize `LibSSL` to implement the CN-MvTPM Module. It includes creating and managing the vTPM, establishing the secure channel, etc. We utilize `libtpms` and modified `SWTPM` to create the vTPM. We also implement the communication protocol between the vTPM and the ACVM in the vTPM. We patched about 630 LoC to `SWTPM` to implement the functions including getting and extending vTPM's trust proof, getting ACVM's initialization measurement, extending the measurements to PCRs, encrypting/decrypting TPM commands, and so on.

### C. ACVM

In the ACVM, we use the OVMF with our modifications as the firmware and the Linux kernel with our modifications as the system kernel. To make the OVMF support our communication protocol between vTPM and ACVM, we add about 380 LoC to implement the function of obtaining the session key and encrypting/decrypting the TPM commands. To implement the above function in the Linux kernel as well, we added and modified about 260 LoC in it.

## VI. EVALUATION OF T3CVM

This section first analyzes the security of the T3CVM approach as a demonstration that the approach can well defend against the threats defined in the threat model (introduced in Section III.A). Then we evaluate the performance of the T3CVM approach in terms of the overhead of ACVM launching and booting, and the overhead of executing TPM commands.

### A. Security

We formalize the analysis of T3CVM using `ProVerif` [41], focusing on how T3CVM meets the various security requirements (**SR1-SR7**) defined to address the threats presented in the threat model. Due to space limitations, we move the detailed formal analysis to Appendix A of the supplementary file.

### B. Performance

We evaluate the performance of the T3CVM approach, focusing on the overhead during the boot process and the runtime of the user's ACVM. Since our approach approximates the CoCoTPM, we compare the overhead imposed by T3CVM and CoCoTPM in executing TPM commands.

Our evaluation uses two servers, one as a user-trusted entity running TR-Manager, and the other one as an insecure Cloud Node running the CN-TPMCVM and the ACVM. The user-trusted entity server is equipped with an Intel i5-13600K chip with 32G of RAM. It runs a 64-bit Ubuntu 22.04 kernel version v6.1.0. `OpenSSL` 1.1.1q and `libtpms` v0.9.6 are included to support the RoT component. The Cloud Node server is equipped with an AMD EPYC 7763 64-core chip and 256G of RAM. The Cloud Node's system is Ubuntu 22.04 with kernel v6.1.0, which runs `QEMU` version 7.2.0. The CVMs use AMD's recommended software stack[35], with OVMF version edk2-stable202205, and the guest system is Ubuntu 20.04, with kernel version v6.5.0. The above components are modified to support our approach.

*1) ACVM's Boot Time*

Our approach modifies the OVMF as well as the kernel. To fully evaluate the impact of these modifications on the ACVM's boot time, we divide the ACVM's boot process into three parts and evaluate the boot time for each part separately:

- **OVMF's Execution Time** of ACVM.
- **Kernel's Initialization Time** of ACVM.
- **User Space's Initialization Time** of ACVM.

AMD SEV officially recommends using encrypted images combined with CVMs for better data protection. The encrypted images can significantly affect the boot time of CVMs, so we consider three cases:

i). **Ours (with Image Encryption)**, denotes our approach (T3CVM) with the encrypted image.
ii). **Ours**, denotes our approach without encrypted image.
iii). **CVM**, denotes the regular user's SEV-based ACVM.

We take the average value for each case of 1000 times of ACVM launching as a set of experiments. According to the requirements of the T3CVM, our components communicate with each other through the TCP/IP. To minimize the impact of the network, we place the components in a Local Area Network (LAN) and do 5 sets of experiments to measure the average time. TABLE II shows the ACVM's boot time for each phase.

TABLE II
ACVM'S BOOT TIME FOR EACH PHASE

| Phases of ACVM Booting | Ours (with Encrypted Image) | Ours | CVM |
|---|---|---|---|
| OVMF's Execution Time | 26.556s | 12.973s | 5.263s |
| Kernel's Initialization Time | 6.089s | 3.207s | 2.458s |
| User Space's Initialization Time | 13.352s | 13.233s | 13.366s |

*OVMF's Execution Time.* We find that our approach with the encrypted image is about 2.1 times slower than the case without the encrypted image. This is due to the overhead caused by the Grub implementation in OVMF [40] when decrypting the image. Our approach without the encrypted image is about 2.5 times slower than a regular CVM. Our approach adds overhead during OVMF's execution process because we add the functions of obtaining the session key and encrypting/decrypting TPM commands. As seen above, our approach ensures secure communication and adds some overhead.

*Kernel's Initialization Time.* We find that during kernel initialization, our approach with the encrypted image takes 2 times longer than the case without the encrypted image. This is because this case requires decrypting the image during the kernel initialization phase. Our approach without the encrypted image is 1.3 times longer than the regular CVM. This is due to the overhead caused by the CN-CDriver in the kernel encrypting TPM commands. As can be seen, our approach secures the TPM data transfer while incurring very little additional overhead.

*User Space's Initialization Time.* In the phase of user space initialization, the difference between the three cases is very small and remains basically the same. T3CVM does not add extra overhead in the initialization phase in user space because our experiments do not use the IMA module. This shows that our approach does not impose additional overhead on the system after the kernel initialization of ACVM is complete.

TABLE III
TIME FOR ACVM TO EXECUTE TPM COMMANDS

| TPM Cmds | Ours | ACVM+vTPM | ACVM+ CVM (vTPM) |
|---|---|---|---|
| Get Random | 15.0ms | 9.0ms | 9.0ms |
| PCR Read | 20.4ms | 14.2ms | 14.8ms |
| PCR Extend | 10.5ms | 5.9ms | 6.2ms |
| Hash | 10.2ms | 6.1ms | 6.1ms |
| RSA Encrypt | 72.7ms | 45.3ms | 48.9ms |
| RSA Decrypt | 127.4ms | 95.9ms | 89.0ms |
| AES Encrypt | 108.6ms | 73.8ms | 75.3ms |
| AES Decrypt | 108.3ms | 75.6ms | 80.8ms |
| Create | 163.0ms | 126.1ms | 118.5ms |
| Sign | 162.5ms | 117.9ms | 122.2ms |
| Verify Signature | 87.2ms | 61.5ms | 67.1ms |

*2) Performance of TPM Commands*

T3CVM implements vTPM in a CVM and adds a secure communication protocol. We want to evaluate the additional overhead added by T3CVM in two aspects:

- Additional Overhead Introduced by Security Protocol.
- Additional Overhead Introduced by Implementing vTPM into CVM.

The execution time of TPM commands is considered in three cases:

i). **Ours**, denotes our approach (T3CVM).
ii). **ACVM+vTPM**, denotes that a regular user's SEV-based ACVM connects with `SWTPM` as a backend to a vTPM via TCP/IP.
iii). **ACVM+CVM (vTPM)**, denotes a regular user's SEV-based ACVM connects with a vTPM running in a CVM via TCP/IP.

We evaluate the execution time of the TPM commands by invoking various TPM commands using the `tpm2-tools` package [42] from the user space of the ACVM. We can evaluate the overhead imposed by the implementation of vTPM in CVM through ACVM+CVM (vTPM) vs. ACVM+vTPM. Also, the comparison of Ours vs. ACVM+CVM (vTPM) can evaluate the overhead imposed by the secure communication protocols implemented in our approach. We do not compare the performance of T3CVM with the physical TPM. This is because the experiments in [24] and [26] can demonstrate that software-based implementations of vTPMs outperform physical TPM in most cases. We test each TPM command 1000 times to get the average time, conduct 3 experiments, and average them to try to mitigate the effect of the network on the experiments. TABLE III shows the execution time of the commonly-used TPM commands.

*Additional Overhead Introduced by Security Protocol.* We find that our approach adds about 45% additional overhead on average for TPM command execution compared to ACVM+CVM (vTPM). This is because the secure communication protocol protects the privacy of each TPM command using the AEAD algorithm. As a result, additional encryption/decryption operations and the increased length of each command also impose some communication overhead. Further, T3CVM adds about 67% overhead in small data-volume TPM commands such as GetRandom, Hash, and PCR Extend and about 40% in large data-volume commands such as RSA Decrypt, Create, and Sign. This is because the communication overhead is the main reason for the performance impact when executing the large data-volume commands. Therefore, the T3CVM's security protocol introduces less additional overhead.

*Additional Overhead Introduced by Implementing vTPM in CVM.* We find that ACVM+CVM (vTPM) imposes less than 10% additional overhead on the execution time of all TPM commands compared to ACVM+vTPM. As seen, the implementation of vTPM in a CVM has a small impact on the execution time of TPM commands. Notably, the ACVM+CVM (vTPM) takes less time than the ACVM+vTPM when executing the RSA Decrypt and Create commands. We evaluate the data volume transferred by the commands and find that the RSA Decrypt and Create commands transmit the largest amount of data. Because of the effects of system scheduling and data transfer, vTPM running in a CVM may have less performance overhead when executing TPM commands with large data volumes. We find that our approach brings less extra overhead when the ACVM needs to transmit a large amount of data.

In summary, T3CVM imposes some additional overhead in executing TPM commands. This overhead is mainly caused by the privacy protection of TPM commands by secure communication protocols. The implementation of vTPM in CVM imposes less extra overhead and even has some advantages when transferring TPM commands with large data volumes.

Since our approach is similar to CoCoTPM [26], we also compared the performance of T3CVM with CoCoTPM in executing TPM commands. Since the source code of the CoCoTPM is not available, we choose the experimental data in their paper to compare with the approximate configuration of our approach. We choose two experimental cases from CoCoTPM to calculate the difference to represent the overhead introduced by CoCoTPM, and calculate the difference in our approach. The experiment results are shown in TABLE IV.

TABLE IV
COMPARISON OF T3CVM AND CoCoTPM EXECUTION TIME FOR TPM COMMANDS

| TPM Cmds | CoCoTPM's Overhead* | Ours Overhead |
|---|---|---|
| **Get Random** | 7.0ms | 7.0ms |
| **PCR Read** | 11.0ms | 6.2ms |
| **Hash** | 4.0ms | 4.1ms |
| **Create** | 58.0ms | 36.9ms |

\* Experimental data from [26].

## VII. CONCLUSION

This paper aims for an approach to build a complete trust chain with a well-defined root-of-trust and then use it to generate as a trust proof of the AMD SEV-based CVM running in insecure cloud nodes. Our proposed approach, T3CVM, tackles the existing approach's weaknesses including lack of a well-defined root-of-trust, lack of vTPM protection, and lack of vTPM's trust proof. We formally analyze the security attributes of the T3CVM, showing that our approach addresses the above issues. We also implement our approach and the experiment results show that T3CVM can establish secure CVMs at a small cost. Our approach improves the security of CVM with less overhead.